\documentclass[11pt,a4paper]{article}

\usepackage[utf8]{inputenc}
\usepackage{doi}
\usepackage{amsmath}
\usepackage{hyperref}
\usepackage{graphicx}
\usepackage{cite}
\usepackage{units}
\usepackage[bitstream-charter]{mathdesign}

\begin{document}

\begin{center}{\Large \textbf{
Study of the properties of sustainable materials for shielding against ionizing radiations\\
}}\end{center}

\begin{center}\textbf{
Julien Faivre\textsuperscript{1$\star$}
}\end{center}

\begin{center}
{\bf 1} Laboratoire de Physique Subatomique et de Cosmologie, Universit\'e Grenoble-Alpes, CNRS-IN$_{2}$P$_{3}$, Grenoble, France
\\[\baselineskip]
$\star$ \href{mailto:julien.faivre@lpsc.in2p3.fr}{\small julien.faivre@lpsc.in2p3.fr}
\end{center}

\section*{Abstract}
\textbf{\boldmath{%
The \textsc{SuShi} project aims at determining how sustainable materials can be used to shield against ionizing radiations. This contribution presents a comparison of the effect of concrete and rammed raw earth on photons. For energies around the \unit{MeV}, where interactions by Compton scattering dominate, changes in the material composition have virtually no effect. The density difference results in \unit[20]{\%} more thickness needed for earth to achieve the same flux attenuation as concrete. The same value is found for higher energies, where photons produce electromagnetic showers. Below \unit[0.1]{MeV}, material composition and earth moisture content matter but the attenuation is much larger.
}}

\vspace{\baselineskip}

\section{Introduction}
\label{sec:intro}

For new projects in high-energy physics, or extensions of existing installations, concrete is a sizeable source of greenhouse gases emissions~\cite{bib:arXiv:sustainabilityFutureAccel,bib:article:FCCfeasibilityVol3} and of waste~\cite{bib:report:EEA2025,bib:article:constructionDemolitionWasteNLworld,bib:article:dechetsConstructionParis}. While concrete can not be easily avoided for elements like the tunnel vaults, the concrete blocks used for shielding against ionizing radiations could instead be made of more sustainable materials, which generate fewer emissions, consume less embodied energy, and can be recycled~\cite{bib:article:SoudaniIntroTerre,bib:article:IndiaEarthBetterSustThanConcrete,bib:article:reviewEarthBetterSustThanConcrete}.

Shieldings containing a sustainable material already exist~\cite{bib:bookChapter:MedAustron,bib:site:ForsterPatent,bib:article:sandwichSEEIIST} but not under the form of concrete-free movable blocks. {\sc SuShi} ({\sc Su}stainable {\sc Shi}eldings) is a project which general scope is to determine in which conditions such materials can be used as shielding. This contribution focusses on comparing the effect of raw earth and concrete on photons.

The drivers of the effects of matter on an ionizing radiation are the atomic densities of the various elements, which must therefore first be determined by chosing their mass fractions and the material density.

\section{Composition of the studied materials}
\label{sec:composition}

In common language, ``concrete'' refers to a blending of cement, obtained by heating a mixture mainly composed of clays and limestone, with aggregates, i.e. sand and gravels. The nominal composition used in this contribution relies on the density and mass fractions of the various elements from~\cite{bib:site:NISTcementPortlandComposition}. Other sets of mass fractions were also estimated by using the proportion and composition of cement of~\cite{bib:article:cementCompositionAndConcreteMix} and varying the composition of the aggregates. These were chosen siliceous by default, by using the composition of granite OM\_3 in~\cite{bib:article:graniteComposition}, but a calcareous composition was also tested by using the high-purity limestone sample LS3 in~\cite{bib:article:limestoneComposition}.

Rammed raw earth~\cite{bib:book:rammedEarthTechnique,bib:book:GernotMinke} is a sustainable material widely used in construction, which can take the form of prefabricated blocks. Raw earths which are suitable for building naturally contain sand and gravels, as well as three-layer non-swelling clays like illite~\cite{bib:phd:AgostinoWalterBruno2016}, which act as a binder. The similar composition with respect to concrete and the only $\simeq \unit[20]{\%}$ lower density of rammed earth seem to make it a good candidate to replace concrete.

The same aggregate compositions as for concrete were used. As per clay, suitable mass fractions for the rammed earth technique are most often found in the range 10-\unit[20]{\%}~\cite{bib:article:clayContent2,bib:article:SoudaniThermique,bib:article:analysisRammedEarthPortugal}. A \unit[10]{\%} clay content was used for the nominal composition, but other mass fractions were obtained by setting it to \unit[20]{\%}.

The clay composition was chosen to be \unit[100]{\%} illite, which general formula writes K$_{x+z}$ (Mg$_x$Al$_{2-x}$)(Al$_z$Si$_{4-z}$)O$_{10}$(OH)$_2$. The amount $x+z \in [0.6 ; 0.85]$ of potassium~\cite{bib:site:mineralDBBRGM,bib:site:mineralDBMindat}, occasionally replaced by hydronium~\cite{bib:site:mineralDBMindat}, is such that the electric charge is balanced. In the first parenthesis, a small fraction of Mg$^{2+}$ and Al$^{3+}$ ions can be respectively replaced by Fe$^{2+}$ and Fe$^{3+}$, thereby changing the photon attenuation owing to the larger atomic number of iron.

This preliminary study was done with unphysical but simpler compositions K(Mg$_x$Al$_{2-x-y}$ Fe$_y$)(Al$_z$Si$_{4-z}$)O$_{10}$(OH)$_2$, where $(x,y,z)$ were chosen as $(0,0,0)$ for the nominal composition, leading to atomic fractions close to those given in~\cite{bib:article:illiteCompositionCZ}, and as $(2,0,4)$ (lowest atomic numbers) and $(0,2,0)$ (highest) for the most extreme cases.

For both materials, the same density was taken for their various element compositions: $\unit[2.3]{g/cm^3}$ for concrete, and $\unit[1.9]{g/cm^3}$ for dry rammed earth.

Finally, rammed earth is a hygroscopic and porous material which absorbs or desorbs moisture to maintain equilibrium with the surrounding air, and therefore contains a mass fraction of water varying in the range 0.5-\unit[3]{\%}~\cite{bib:phd:LucileSoudani2016,bib:phd:LongfeiXu2018}. A fraction of \unit[2]{\%} was chosen as a nominal amount, and added to the atom contents.

\section{Interaction of photons in the MeV range}
\label{sec:photonsMeV}

When crossing matter, the interactions of photons of a given energy between the \unit{keV} and a few \unit{MeV} are described by the exponential attenuation of the initial photon flux $\Phi_0$ as a function of the crossed material thickness $x$, written as $\Phi(x) = \Phi_0 e^{-\mu x}$~\cite{bib:book:Carron}.

For compound materials, the linear attenuation coefficient can be written as $\mu = \sum_i n_i \sigma_i$, with $n_i$ the atomic densities and $\sigma_i$ the cross sections of each element $i$. The databases provide the mass attenuation coefficient ({\sc mac}) $\frac{\mu}{\rho} = \frac{\mathscr{N}_A}{M} \sigma$, where $\rho$ is the density, $M$ the molar mass and $\mathscr{N}_A$ the Avogadro number. The flux attenuation in a compound material denoted $c$ can therefore be calculated as:
\begin{equation}
\frac{\Phi(x)}{\Phi_0} = e^{- \left(\frac{\mu}{\rho}\right)_{c} \rho_{c} x } \qquad\text{where}\quad \left(\frac{\mu}{\rho}\right)_{c} = \sum_i p_{m_i} \left(\frac{\mu}{\rho}\right)_i
\end{equation}
where $p_{m_i}$ is the mass fraction of element $i$ in the compound material.

\begin{figure}
\centering
\begin{minipage}{7.2cm}
\centering
\includegraphics[height=5.5cm]{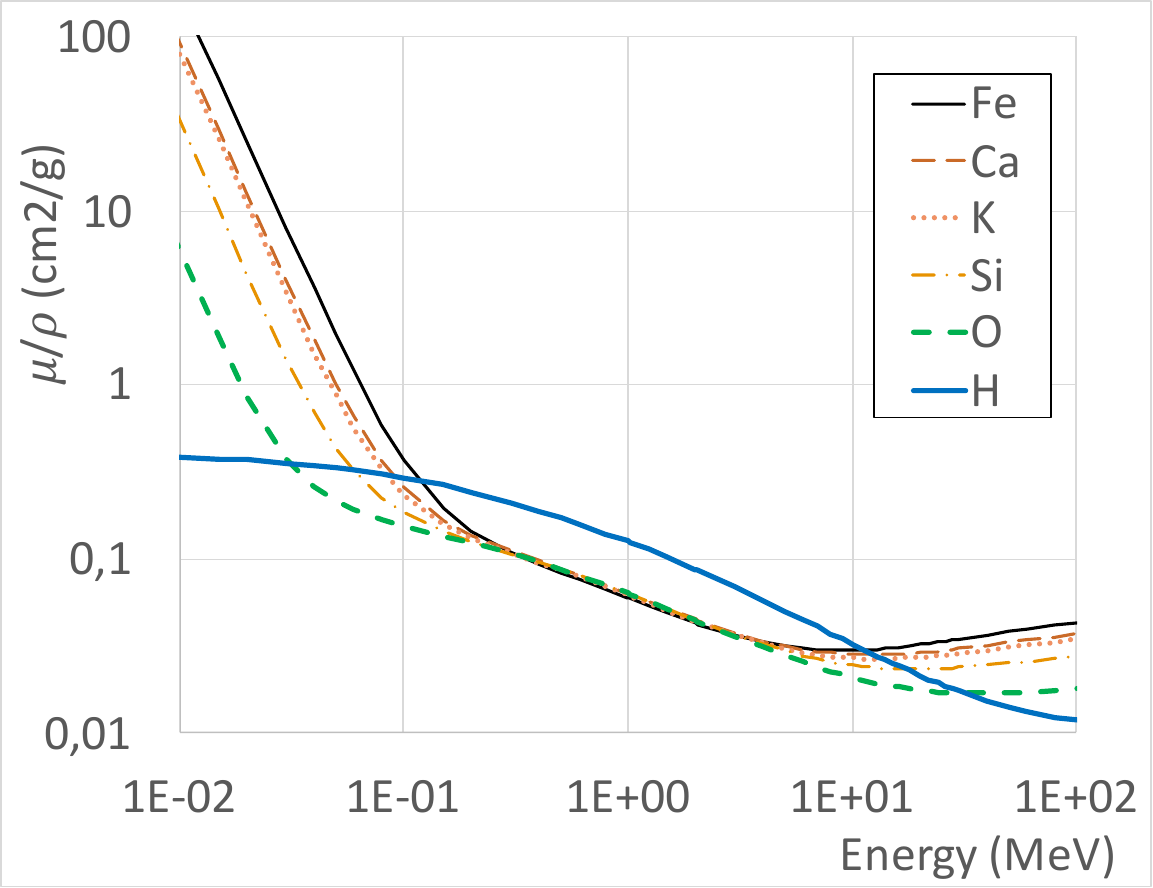}
\caption{Mass attenuation coefficient~\protect\cite{bib:site:XCOM} as a function of the photon energy, for hydrogen (blue thick solid line) and several other elements found in sustainable materials, from oxygen (green dashed line) to iron (black thin solid line).}
\label{fig:muOverRhoVsE:NIST}
\end{minipage}
\hfill
\begin{minipage}{7.2cm}
\centering
\includegraphics[height=5.5cm]{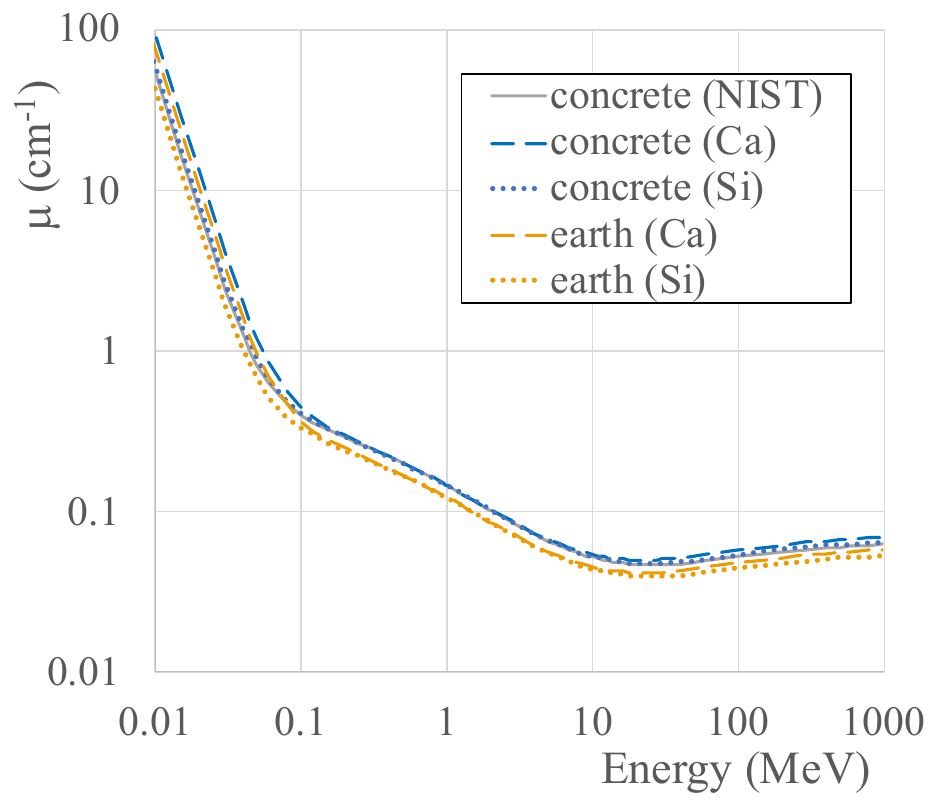}
\caption{(Color online) Energy dependence of linear attenuation coefficients calculated for concrete (blue) and raw earth (orange), for calcareous (dashed) and siliceous (dotted) aggregates, as well as for NIST concrete (solid grey).}
\label{fig:muVsE:ConcreteEarthComposition}
\end{minipage}
\end{figure}

Figure \ref{fig:muOverRhoVsE:NIST} shows the tabulated {\sc mac} versus energy for various elements~\cite{bib:site:XCOM}. Around the \unit{MeV}, photons interact with matter mainly through Compton scattering $\gamma + e^- \rightarrow \gamma + e^-$, which cross section is proportional to the atomic number $Z$, and for which the {\sc mac} is thus roughly proportional to the proton-to-nucleon ratio $Z/A$. In this energy range, the photon flux attenuation in a material is therefore expected to depend only on the Hydrogen content and on the density of this material.

At lower energies, the photoelectric effect $\gamma + X \rightarrow e^- + X^+$ dominates, and the strong dependence of its cross-section on the atomic number is expected to induce a sizable variation of the attenuation coefficient with the material composition.

Figure~\ref{fig:muVsE:ConcreteEarthComposition} shows the linear attenuation coefficient $\mu$ calculated for rammed raw earth and concrete with calcareous and siliceous aggregates, as well as for the nominal concrete. The different material densities result in larger values of $\mu$ for concrete at all energies, such that a rammed earth block would have to be $\simeq \unit[20]{\%}$ thicker than a concrete block to achieve the same flux attenuation.

As expected, it is only below \unit[200]{keV} that the photoelectric effect results in differences between the various material compositions tested. These differences are mainly driven by the composition of the aggregates; they reach a few tens of \%, but the $E_\gamma^{-7/2}$ dependence of the cross-section on the photon energy results anyway in much larger attenuation factors than for energies around the \unit{MeV}.

Adding moisture results in increasing the global atom density of the material and therefore the attenuation factor $e^{-\mu x}$ as well. A \unit[1]{\%} seasonal variation of moisture in a raw earth block was found to have little impact on the attenuation factor of \unit[1]{MeV} photons, but results in a variation of \unit[10]{\%} around \unit[50]{keV}, and of a factor of 2 around \unit[20]{keV}, due to the fast energy dependence of $\mu$.

Finally, rammed earth attenuation coefficients were measured at 45, 344, 779, 964, 1112 and \unit[1408]{keV} by using the $\gamma$ emitted by a $^{152}$Eu source, a Germanium detector, and two \unit[6]{cm} thick blocks of rammed recycled earth, of densities 1.6 and $\unit[2.0]{g/cm^3}$. The values calculated with the nominal raw earth atom composition are in agreement with the data.

\section{Interaction of photons in the GeV range}
\label{sec:photonsGeV}

For energies above tens of \unit{MeV}, the dominant interaction mode of photons is pair creation ($\gamma \rightarrow e^+ + e^-$)~\cite{bib:book:Carron}. The electron and positron created interact by Bremsstrahlung ($e \rightarrow e + \gamma$), i.e. emission of a photon carrying a significant fraction of the electron or positron energy, which can interact again in matter by pair creation.

This electromagnetic shower keeps developping until the average energy of the particles falls below the critical energy $E_c$, which is the energy below which Bremsstrahlung stops being the dominant cause of electron energy loss. $E_c$ is also close to the energy below which Compton scattering dominates over pair creation.

Denoting $X_0$ the radiation length, i.e. the average distance between two Bremsstrahlung emissions, the length of an electromagnetic shower produced by a photon with an initial energy $E_0$ can therefore be calculated. In a compound material, it writes~\cite{bib:book:LeroyRancoita}:
\begin{equation}
L(E_0) = \frac{X_{0,c}}{\ln 2} \ln \left(\frac{E_0}{E_c}\right) \qquad\text{with}\quad \rho_c \; X_{0,c} = \frac{1}{\sum_i \frac{p_{m_i}}{\rho_i X_{0,i}}}
\end{equation}
where $\rho_c$ is the density of the compound material, $p_{m_i}$ the mass fractions of the various elements $i$ in the compound, and $\rho_i \, X_{0,i}$ the product of the density by the radiation length for each element.

The values of $\rho_i \, X_{0,i}$ were taken from a database~\cite{bib:site:PDGdatabase}, as well as the value of $E_c$ for concrete. With the assumption that the critical energy in raw earth has the same value as in concrete, the ratio of the length of the electromagnetic showers in raw earth and in concrete does not depend on the photon energy, and was found to be below 1.2 for all the tested material compositions.

\section{Conclusion}
\label{sec:concl}

The possibility to use sustainable materials as shielding against ionizing radiations was explored by focussing on rammed raw earth, which characteristics resemble that of the widely used concrete blocks. The length of electromagnetic showers created by photons with energies above a fraction of \unit{GeV} were calculated, as well as the linear attenuation coefficient for photons with energies down to a few tens of \unit{keV}, for various material compositions.

In both cases, the results indicate that density is the factor which drives the difference between concrete and raw earth, and that the additionnal earth thickness needed to reach a similar shielding to concrete is about \unit[20]{\%}, in agreement with preliminary measurements.

This encouraging results have to be strengthened by refined studies, as well as by testing other ionizing radiations.

\section*{Acknowledgements}

Part of this work was made during the internship of two undergraduate students: Fran\c{c}ois Viars and Flora Coulaud-Bastien. The author thanks them for their curiosity and fearlessness to get off the beaten track, which lead them to work on this peculiar topic.

\paragraph{Funding information}
No budget was needed to carry out this study.

\bibliography{proceedingsSustHEP2026_Faivre_v3_arXiv.bib}

\end{document}